\documentclass[3p,times]{elsarticle}

\usepackage{ecrc}
\usepackage[table,xcdraw]{xcolor}

\volume{00}

\firstpage{1}

\journalname{Computers and Fluids}

\runauth{A. AlOnazi et al.}

\jid{procs}

\jnltitlelogo{Computers and Fluids}

\CopyrightLine{2014}{Published by Elsevier Ltd.}

\usepackage{amssymb}

\usepackage[figuresright]{rotating}
\usepackage{color}
\usepackage{multirow}

\usepackage{epstopdf}
\usepackage{todonotes}

\begin{document}

\begin{frontmatter}



\dochead{ParCFD 2014}

\title{Design and Optimization of OpenFOAM-based CFD Applications for Hybrid and Heterogeneous HPC Platforms}


\author{Amani AlOnazi$^{*}$, David E. Keyes$^{*}$, Alexey Lastovetsky$^{\dag}$, Vladimir Rychkov$^{\dag}$}

\address{$^{*}$Extreme Computing Research Center, KAUST, Thuwal 23955-6900, Saudi Arabia, \\
e-mail: \{amani.alonazi, david.keyes\}@kaust.edu.sa\\
$^{\dag}$Heterogeneous Computing Laboratory, UCD, Dublin 4, Ireland, \\
e-mail:\{alexey.lastovetsky, vladimir.rychkov\}@ucd.ie}

\begin{abstract}
Hardware-aware design and optimization is crucial in exploiting emerging architectures for PDE-based computational fluid dynamics applications. In this work, we study optimizations aimed at acceleration of OpenFOAM-based applications on emerging hybrid heterogeneous platforms. OpenFOAM uses MPI to provide parallel multi-processor functionality, which scales well on homogeneous systems but does not fully utilize the potential per-node performance on hybrid heterogeneous platforms. In our study, we use two OpenFOAM applications, icoFoam and laplacianFoam, both based on Krylov iterative methods. We propose a number of optimizations of the dominant kernel of the Krylov solver, aimed at acceleration of the overall execution of the applications on modern GPU-accelerated heterogeneous platforms. Experimental results show that the proposed hybrid implementation significantly outperforms the state-of-the-art implementation.

\end{abstract}


\end{frontmatter}


\section{INTRODUCTION}
\label{INTRODUCTION}

Current HPC systems are complex and difficult to utilize and manage at their extremes. The heterogeneity of these platforms leads to several challenges and much contemporary attention is devoted to new software solutions. However, CFD packages, until recently, have been aimed at homogeneous systems and the parallel approach is mostly based on the process-level Message Passing Interface (MPI). This trend in the HPC platforms invites redesign of the CFD packages or the algorithms themselves to use these platforms efficiently.\\

This paper proposes optimizations aimed at accelerating numerical simulations on modern and emerging hybrid heterogeneous platforms, which are illustrated in two OpenFOAM \cite{openfoam} applications: icoFoam and laplacianFoam, which are representative of many low-order discretization of incompressible flow. We selected the OpenFOAM CFD package because of its popularity as an open source library with broad adoption. Both icoFoam, which is an incompressible flow solver, and laplacianFoam, which solves the Poisson equation, for e.g., thermal diffusion, are based on Krylov iterative methods such as conjugate gradient method. \\

\begin{figure}[h]
\begin{center}
\emph{$A)$}\includegraphics[width=0.47\linewidth,height=0.25\textheight]{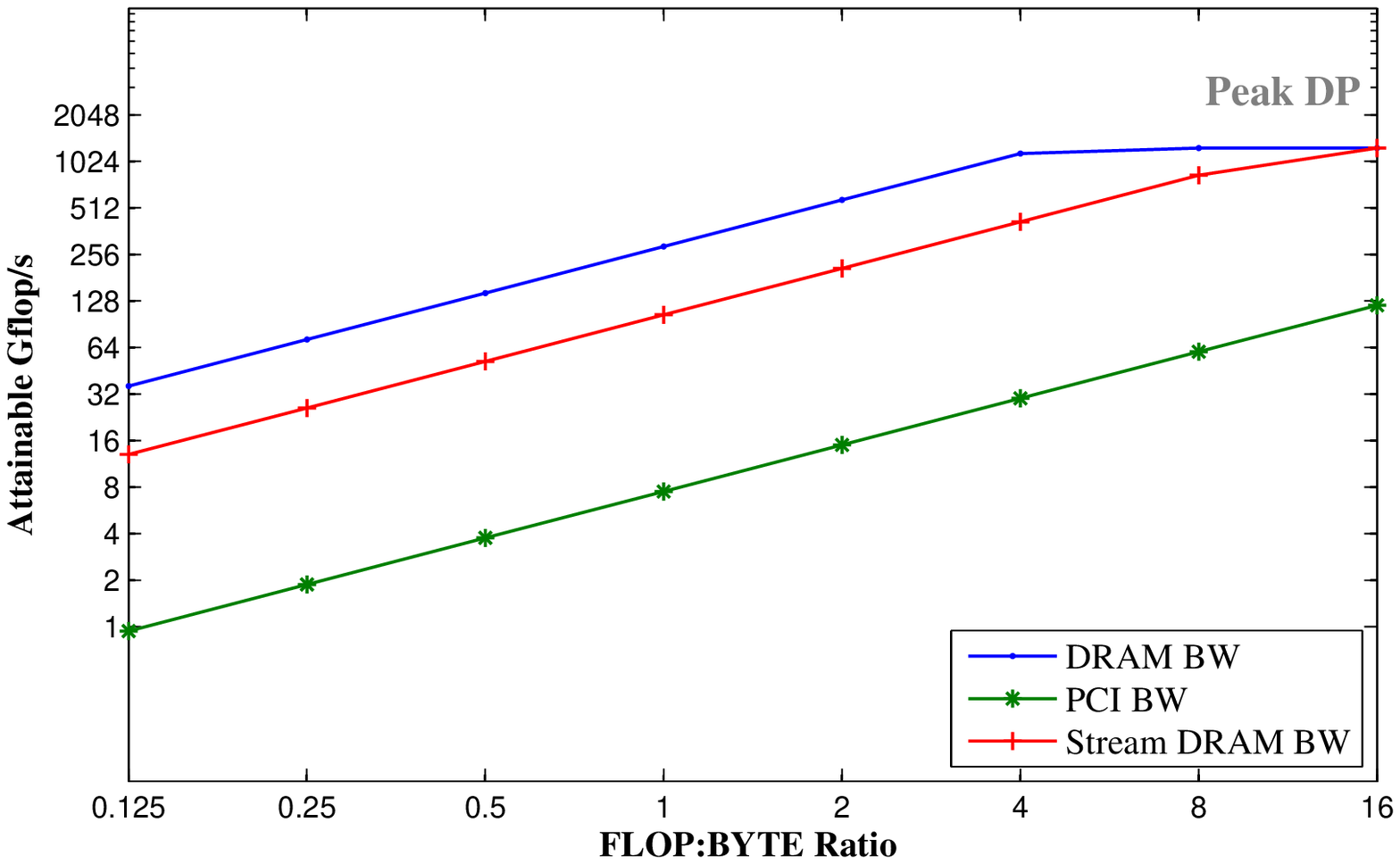}
\emph{$B)$}\includegraphics[width=0.47\linewidth,height=0.25\textheight]{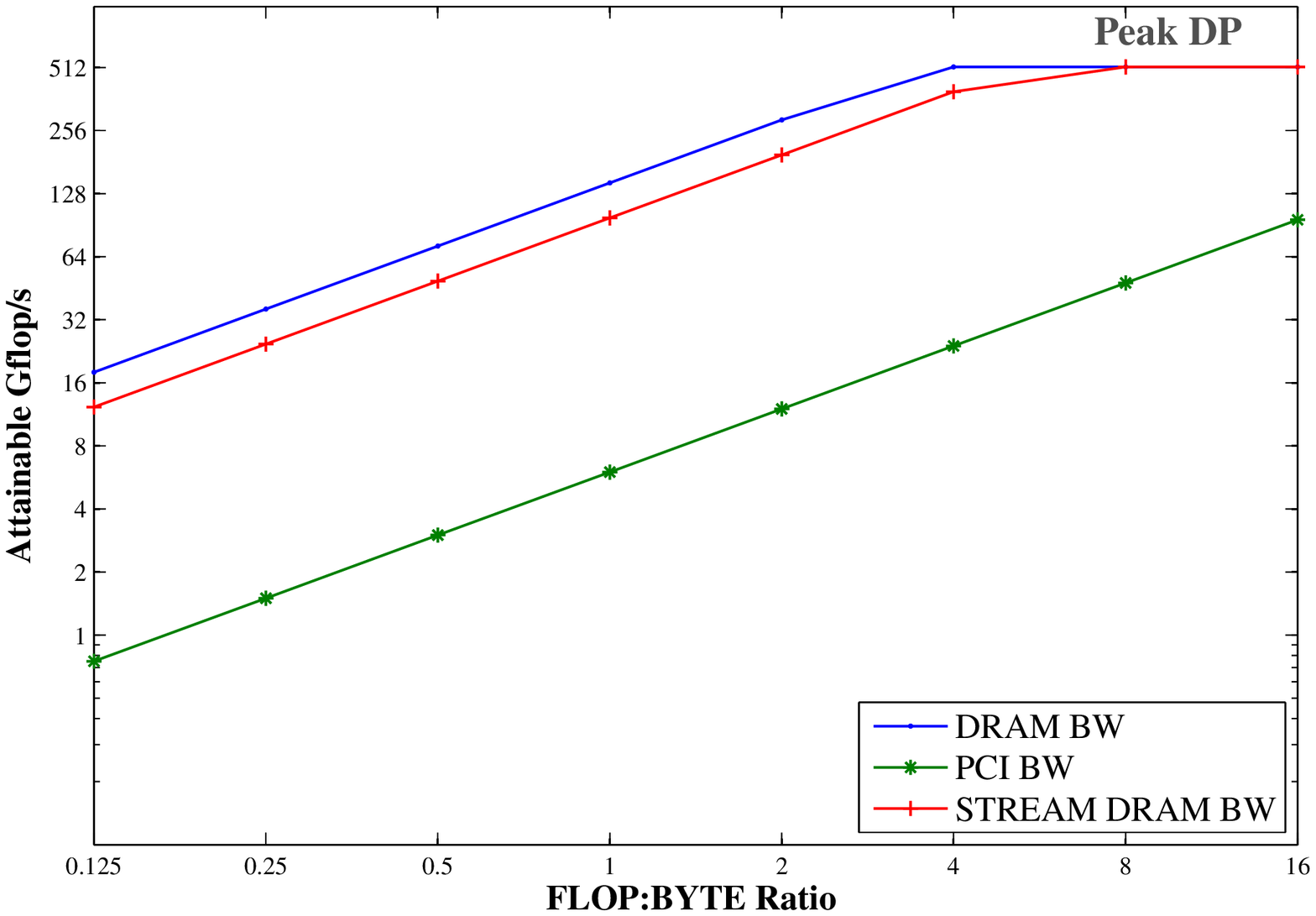}
\caption{The attainable performance on: \emph{$A)$} a GeForce GTX Titan, \emph{$B)$} a Tesla C2050 (Fermi)  node, according to the roofline model combined with the low memory bandwidth of the device measured by stream benchmark.}
\label{perf1}
\end{center}
\end{figure}

The behavior of the conjugate gradient method (CG) has been extensively studied. The performance of this method per iteration depends on the performance of each individual kernel. There are three primary kernels in CG: vector inner products, vector-vector updates, and sparse matrix-vector multiplication. The parallel-run of the conjugate gradient on a distributed memory architecture using MPI parallelizes the computations and introduces communications as well. Two types of communications are necessary to support the algorithm. The first type is point-to-point communication during the matrix-vector multiplication in order to include the influence of the parallel interfaces with neighbor processors. The second one is a collective communication to reduce the sum of scalars as part of forming inner products. It occurs three times while calculating global variable and updating the residual norm. Figure \ref{perf1} illustrates the roofline model \cite{williams2009roofline} of both Fermi and Titan GPUs platform, which shows the attainable performance when it is limited by either device memory or PCIe bandwidth. The stream memory bandwidth is measured using our implementation of the STREAM benchmark \cite{streambench}. Utilizing such heterogeneous platform for memory bound kernel, CG for instance, leads to two main challenges: data movements and workload balancing across hybrid heterogeneous devices. \\

The proposed optimizations are aimed at accelerating the icoFoam and laplacianFoam solvers on heterogeneous hybrid nodes beyond their current performance. A hybrid conjugate gradient solver is designed and implemented combining MPI and CUDA routines. A load-balancing step is applied using heterogeneous decomposition, which decomposes the computations taking into account the performance of each computing device and minimizing communication. The heterogeneous decomposition method consists of two steps: building the accurate performance model of the application using the approach of FuPerMod \cite{fupermode, fupermode2, fupermode3, fpm_14}, and then using this performance model as input to MeTiS \cite{MeTiS} /SCOTCH \cite{scotch} libraries. In addition, we present a new pipelined conjugate gradient solver as an algorithmic improvement inspired by the recent work of Ghysels and Vanroose \cite{pipecg} and implemented using MPI, CUDA, and a hybrid technique. \\

This paper is organized as follows. Section 2 provides a brief introduction to OpenFOAM and the applications selected for this study. Section 3 outlines related work. The proposed optimizations are presented in Section 4. Section 5 gives experimental evaluation of the performance of our optimizations. Section 6 concludes the paper.

\section{OpenFOAM}
\label{OpenFOAM}
Open source Field Operation And Manipulation (OpenFOAM) \cite{openfoam} is a library written in C++ used to solve partial differential equations (PDEs). It features a wide range of solvers employed in CFD, such as Laplace and Poisson kernels, incompressible flow, multiphase flow, and user-defined models. OpenFOAM uses MPI to provide parallel multi-processor functionality, which scales well on homogenous systems but does not fully utilize potential per-node performance on hybrid heterogeneous platforms. ~\\

The OpenFOAM main library provides a solution framework, including mesh handling, finite volume discretization method, linear system solvers, data structure and input and output handling. An attraction of OpenFOAM is its modularity, which leads to a flexible design by using C++ object oriented structures. One of the modules consists of the linear algebra kernels; this is instrumental in almost all the PDE solver codes. This work selects two applications from the OpenFOAM standard applications set. The first is icoFoam, which solves the incompressible laminar Navier-Stokes equations. It applies the Pressure Implicit with Splitting of Operators (PISO) \cite{piso} algorithm in a time stepping loop. The governing equations are the incompressible continuity equation (\ref{first}) and momentum equation (\ref{second}). 
\begin{equation}\label{first}
\nabla\centerdot u = 0 , 
\end{equation}
\begin{equation}\label{second}
 \frac{\partial u}{\partial t} + \nabla\centerdot (uu) - \nabla\centerdot (\nu \nabla u) = - \nabla p. 
\end{equation}
Here \emph{$u$} is the velocity vector, \emph{$\nu$} is the kinematic viscosity, and \emph{$p$} is the pressure. The PISO algorithm consists of three main stages. The first stage is the momentum predictor, which solves the momentum equation by using an initial or previous pressure field. This solution of the momentum equation gives the velocity field that is not divergence free but approximately satisfies the momentum equation. The second stage is the pressure solution. Third is the explicit velocity correction stage. The velocity field is corrected as a consequence of the pressure distribution. The velocity correction is performed in an explicit manner. The Conjugate Gradient (CG) and BiConjugate Gradient Stabilized (BiCGSTAB) linear solvers are used to solve the pressure and velocity fields respectively.\\

The second selected application is laplacianFoam, which is used to find the solution of the Laplacian equation. The equation contains one variable, a passive scalar, for instance, a temperature, \emph{T}. The matrix is formed and computed as its time-implicit Helmhpltz form, as in \cite{openfoam}, for:
\begin{equation}\label{eighth}
\frac{\partial T}{\partial t} - \nabla (D_{T}\cdot T) = 0 .
\end{equation}
The linear solver that is used to solve the temperature in laplacianFoam is CG \cite{saad_book}, which is one of the best-known Krylov subspace methods. CG is used to solve a system of linear equations in the form:
\begin{equation}\label{nine}
Ax = b .
\end{equation}
Here \emph{$x$} is an unknown vector of size the number of cells in the computational domain, \emph{$b$} is of corresponding length, and \emph{$A$} is a given square, symmetric, positive-definite or semidefinite matrix. The main building blocks of a CG iteration, as formerly mentioned, are: a sparse matrix vector multiply kernel, an optional preconditioning,  three global dot products, and vector scalings and vector-vector additions. 

\subsection{Parallel Approach} 
A common approach for parallelizing partial differential equations is domain decomposition. The mesh and its associated fields are divided into subdomains and allocated to separate processors. OpenFOAM applies domain decomposition to enable process-level parallelism \cite{openfoam}. Thus, the program can run in parallel on separate subdomains, with communications between processors using the MPI communication protocol. The number of subdomains should be equal to (or greater than) the number of processors. Parallel execution engages communication and synchronization between the processors. There are two types of communications: communications between processors hosting the neighboring subdomains, and global communication involving all the processors. As with most mesh-based applications, the communication scheme between processors is based on the halo-layer approach with overlapping elements, which duplicates the cells next to a processor boundary (internal boundary). The halo-layer covers all internal boundaries and is explicitly updated through communications between processors. \\

The most time consuming parts are the sparse linear algebraic kernels. Parallel CG engages point-to-point communication between neighboring processors during the sparse matrix vector kernel and global synchronization for every dot product. 

\section{RELATED WORK}
In the past decade, a number of parallel packages designed for CFD have been proposed including OpenFOAM \cite{openfoam}. Many of these packages are aimed at homogeneous systems, use MPI-based bulk synchronous processing parallelization, and pay no special attention to the underlying hardware. The addition of GPUs to high-end scientific computer systems anticipates new achievable performance levels. As a result, several works attempt to employ GPGPUs in CFD codes and much more attention is focused on sparse linear algebra kernels, such as implementing the conjugate gradient on GPU \cite{ament2010parallel}. These works include the Cufflink library \cite{Cufflink}, which extends the OpenFOAM linear solvers capabilities to perform on GPUs. \\

Pioneering CFD works have addressed heterogeneous and hybrid systems. A parallel simulation of oil extraction has been designed and optimized for heterogeneous networks of computers \cite{alexey98} by applying the performance modeling of the target platform and model-based domain partitioning. Gropp \cite{gropp2001high} studied the hybrid MPI+OpenMP programming model on unstructured implicit CFD solvers and its affects on obtaining per-processor efficiency and parallel efficiency. Accordingly, there have been some attempts to obtain a hybrid parallelization in CFD on multi-core platforms \cite{dagnaaevaluation} such as the CFD solver TAU \cite{tauSolver} that provides hybrid MPI+OpenMP parallelization. There are matrix and graph partitioning libraries, which consider heterogeneity of the devices, such as MeTiS \cite{MeTiS} and SCOTCH \cite{scotch}. These two libraries assume a given vector of positive constants, representing the relative volume of computations to be performed by each processor, to partition a given domain into subdomains, assuming one-to-one mapping between the subdomains and the processors. 
~\\

Recently, there have been attempts to investigate the parallel model of hybrid MPI+GPGPU. Papadrakakis \cite{Papadrakakis20111490}, attempts to balance the computation across heterogeneous hybrid CPU+GPU system. The balancing is performed at runtime by using task-based parallelism and migrations between the compute devices. On the other hand, a general framework is supported by StarPU \cite{starpu} project, which provides and supports task-based programming and scheduling the provided tasks on heterogeneous platform, which combines CPUs and GPUs. The OP2 project \cite{OP2} provides a framework to implement CFD applications using unstructured meshes on different computational hardware including multi-cores and GPUs (many-cores) systems. The parallelization scheme is similar to the task-based parallelism in the way it handles the data dependencies. However, it is not applicable to multi-block codes as is the case in OpenFOAM and there is no special attention to the heterogeneity in the domain decomposition method, which is performed at the MPI level. Moreover, it does not yet extend to multiple GPUs, which is an important direction as the communication architecture of GPUs evolves, ultimately bypassing the PCI-express (PCIe) protocol. 

\section{PROPOSED OPTIMIZATIONS}
The main motivation of this work is to increase the performance of the selected applications and maintain an effective use of the allocated resources. The first solution algorithm proposed here is the hybrid conjugate gradient, which solves the system of linear equations derived from the partial differential equations using the original conjugate gradient method on heterogeneous hybrid platform efficiently. Second, algorithmic improvements of the CG solver that are inspired by the recent work of Ghysels and Vanroose \cite{pipecg}, which introduces more computational steps unnecessary in the original CG while reducing the global communication into one per loop body instead of three, have been implemented in three parallel versions: CUDA, MPI, and hybrid MPI+CUDA. Given the trends of computer architecture, this is a very useful option, as we demonstrate. Third, a load-balancing step is applied using heterogeneous decomposition, which decomposes the computations taking into account the performance of each computing device and minimizing communication. The heterogeneous decomposition combines the performance model and the MeTiS/SCOTCH libraries. \\

\subsection{Hybrid CG and Hybrid Pipelined CG}
The main idea is to distribute the computations across heterogeneous devices, some of them being GPUs, and run these computations in parallel. All the processors employ a process-level parallelism using MPI and each MPI-process, which is connected to a GPU, accelerates the computation using CUDA. Those processors, which are not connected to a GPU run the computation locally. We use the CUSP \cite{cusp} and Thrust \cite{thrust} libraries for implementing CUDA kernels. The computational flow on the CPU and on the GPU starts by decomposing the computational domain into \emph{p} subdomains using OpenFOAM \emph{decomposePar} function, which invokes the domain decomposition method to partition the mesh to a given number of subdomains. The application repeatedly invokes the linear solver, which is, in our case, the hybrid CG. The communication scheme of one iteration requires three global communications and one point-to-point communication between each pair of neighboring subdomains. The matrix is derived from the discretized equation, sparse, and of the dimension of the number of finite volumes. In case of a decomposed computational domain, the local matrix row dimension is number of cells within the subdomain. The hybrid CG solver supports common sparse matrix storage format that is compressed sparse row (CSR). \\

We implement the pipelined conjugate gradient algorithm of Ghysels and VanRoose \cite{pipecg}. It can be considered as a communication startup-reducing algorithmic improvement of the original method. In terms of communication, the main kernels in CG algorithm are sparse matrix-vector multiplication, which requires point-to-point communication among neighboring subdomains, and three dot products, which involve global communication and require participation of all subdomains. The algorithm modifies and reorders the s-step CG method, which is introduced by Chronopoulos and Gear in their original paper \cite{s_step} and minimizes the global communication to only one collective communication per loop body instead of three where the global synchronization can be overlapped by the sparse matrix-vector multiplication kernel instead of updating a vector. Given that the runtime of a vector update is not enough to hide the latency of the global communication, pipelining CG will enhance the performance. Basically, it offers opportunity of better scalability at the price of extra computations, which are relatively cheaper. \\

The hybrid pipelined CG combines the MPI pipelined CG and CUDA pipelined CG kernels. The first communication inside the loop is a single reduction followed by matrix-vector multiplication, which computes the local computations and requires talking with neighboring subdomains to update the processor boundary. The global reduction can be hidden using the asynchronous global communication. However, the point-to-point communication can be between two hosts, two CPUs kernel, or hybrid. These communications require data movements between host and device. All solvers are integrated into OpenFOAM as a dynamic library. \\ 
\begin{figure}[t]
\begin{center}
\includegraphics[width=0.4\linewidth,height=0.27\textheight]{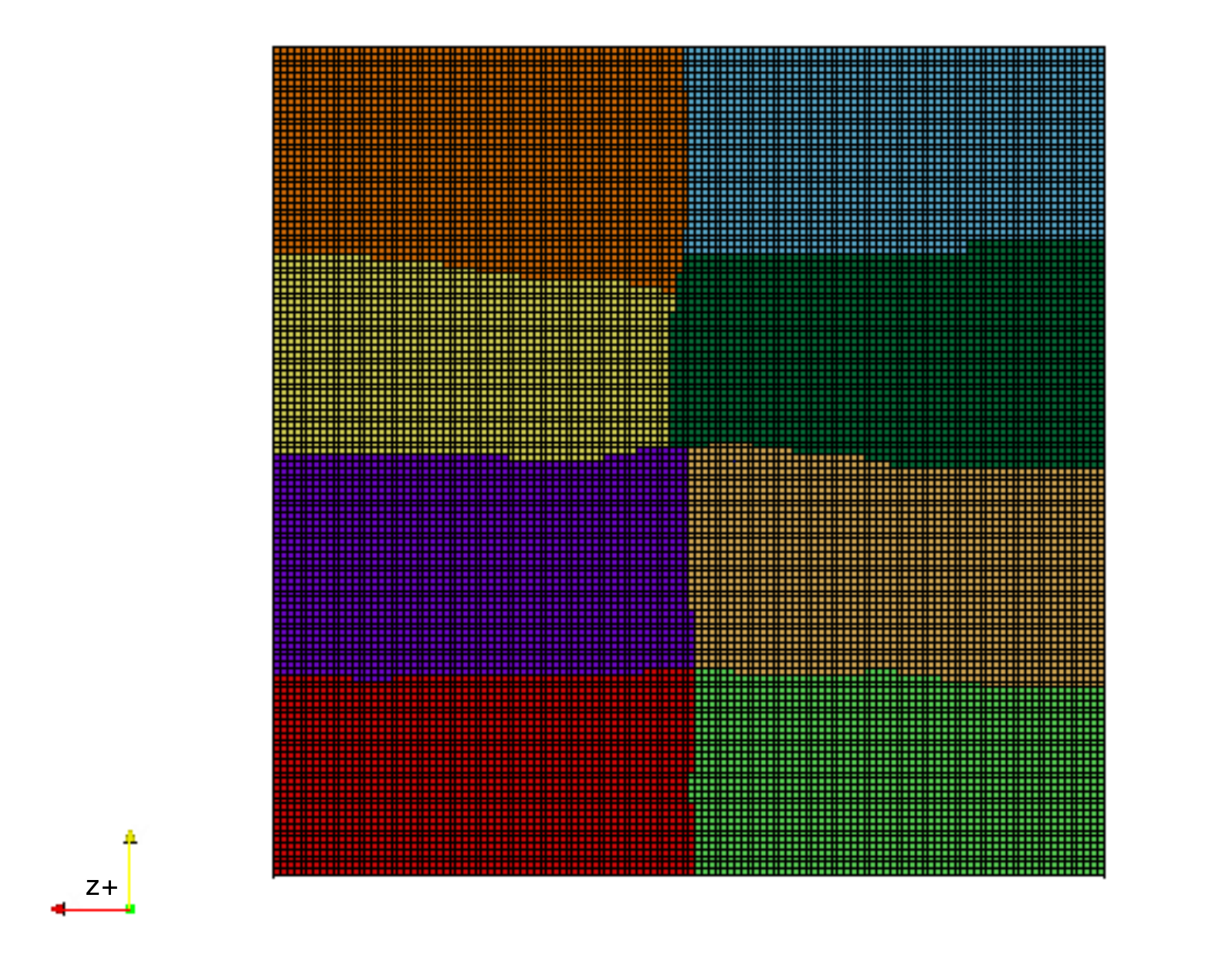} 
\caption{ A $128 \times 128 \times 128$ 3D mesh evenly decomposed using MeTis into 16 subdomains, the view is the face at the positive z-axis.}
\label{pic_evenDecom}
\end{center}
\end{figure}

\begin{figure}[h]
\begin{center}
\includegraphics[width=0.4\linewidth,height=0.27\textheight]{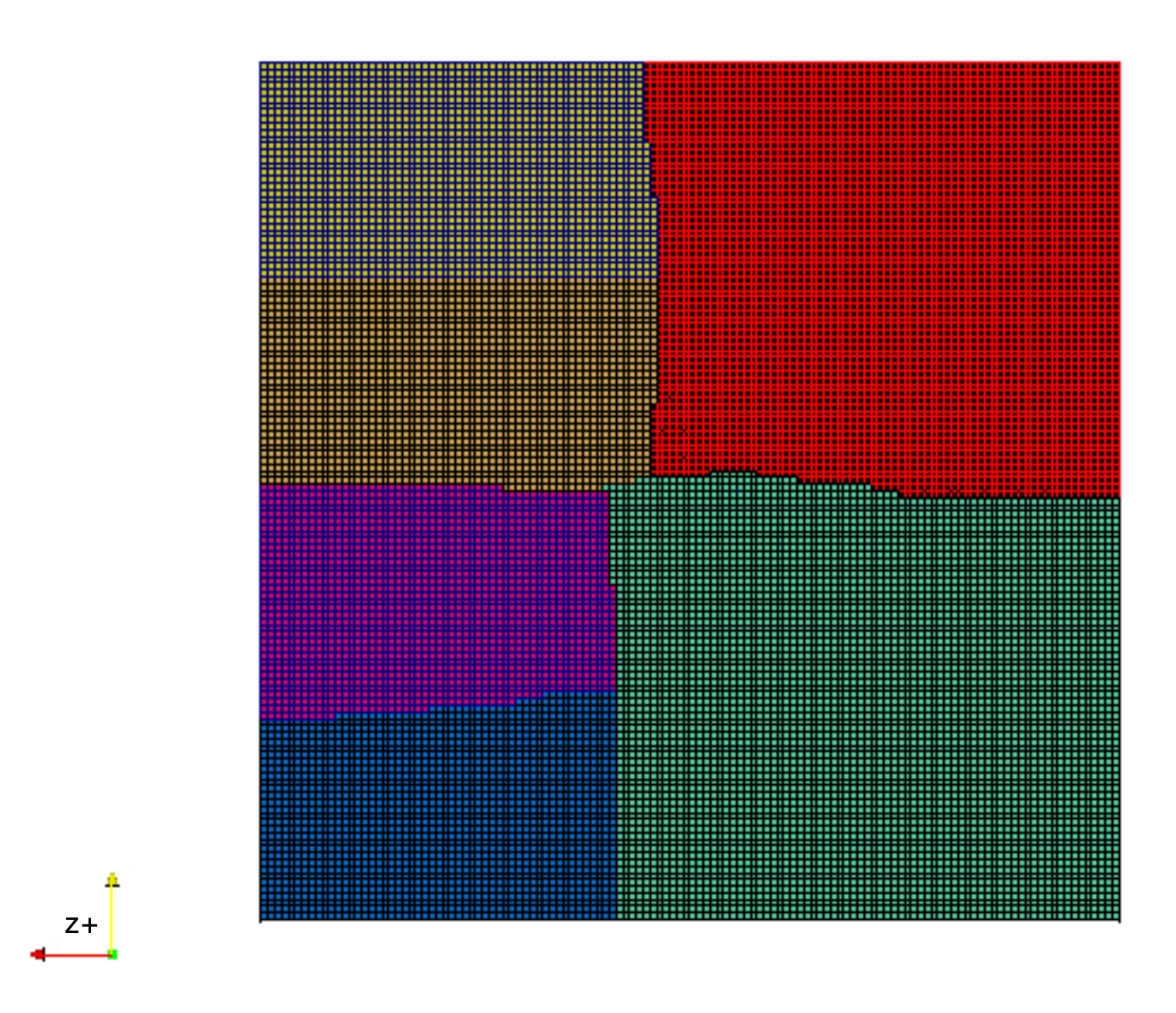} 
\caption{ A $128 \times 128 \times 128$ 3D mesh decomposed using heterogeneous decomposition into 16 subdomains to be allocated to a machine of 16 cores and 2 GPUs, the view is the face at the positive z-axis.}
\label{pic_hetroDecom}
\end{center}
\end{figure}

\subsection{Heterogeneous Decomposition}

The shape and balance of the domain decomposition play an important role in the parallel run. If the domain is not decomposed properly to minimize the number and size of the boundaries between pairs of processors, the parallel run will suffer from an unbalanced workload execution and communications overhead.  The main idea behind the heterogeneous decomposition is to adequately partition and assign the subdomains to the processors in proportion to their performance. Also, the technique should take into account minimizing the communications, so as to minimize inter-processor communications and network congestion. As described in previous sections, OpenFOAM provides static load balancing during the decomposition process through some third-party library, e.g., MeTiS or SCOTCH, prior to running the program. MeTiS and SCOTCH both allow the user to provide a constant number per processor that represents the percentage of data to be computed by this processor. However, neither library provides any methods or algorithms to compute this percentage for balanced workload. It is up to the user to supply these indicators.  \\

In a nutshell, the heterogeneous decomposition combines the performance model \cite{fupermode, fpm_14} and the MeTiS/SCOTCH library. It estimates the relative speed of each compute device in the target platform by running the computational kernel on evenly distributed subdomains, and measures the execution time per processor. Then, it computes the speed of each processor, and builds the performance model of the application. The output of this step is a vector of numbers representing the computational volume per-processor accordance to its speed. The processor computational load will be balanced when each processor's computational work divided by its relative speed is equal. The work shares are provided to MeTiS or SCOTCH to re-decompose the data. \\

Let us assume the hybrid CG solver is the computational kernel to be optimized using the heterogeneous decomposition on structured grid of \emph{$N$} cells. The input is a square sparse matrix \emph{$A_i$}, which contains \emph{$n_i \times n_i$} elements, where \emph{$n_i$} is number of mesh cells allocated to processor \emph{$p_i$}. The number of non-zeros in \emph{$A_i$} is \emph{$n_{i}+2o_{i}$}, where \emph{$o_i$} is number of off-diagonal elements in the upper-triangle of matrix  \emph{$A_i$}.  It is a half-stencil size, i.e., for 3D Poisson, \emph{$o_i$} is three per cell. Then, let us compute the speed \emph{$s_{i}$} of processor \emph{$p_{i}$} during computing the CG kernel as number of non-zeros \emph{$n_{i}+2o_{i}$} divided by the execution time \emph{$T$} per iteration, which can be formulated in equation (\ref{speed_f1}). The asymptotical time complexity of CG kernel per iteration is proportional to number of non-zeros \cite{saad_book}. 

\begin{equation}\label{speed_f1}
s_{i}(n_{i}) = \frac{n_{i}+2o_{i}}{T(n_{i})} .
\end{equation}

\begin{equation}\label{relative_f2}
r_{i}(n_i) = \frac{s_{i}(n_{i})}{\sum_{p}{s_{i}(n_{i})}} .
\end{equation}

\begin{equation}\label{N_dif}
N = \sum_{p}{n_{i}} .
\end{equation}

\begin{equation}\label{percentage_f1}
n_{i, new} = N *r_{i}(n_{i}) .
\end{equation}

Then, the relative speed \emph{$r_{i}$} is computed according to equation (\ref{relative_f2}). After that, the new assigned number of cells to be allocated at processor \emph{$p_{i}$} is computed using equation (\ref{percentage_f1}). This vector of \emph{p} numbers  will be passed as argument to MeTiS/SCOTCH API to re-decompose the computational domain correspondingly. Figure \ref{pic_evenDecom} illustrates a cubic domain that is decomposed evenly into 16 subdomains. If we apply the heterogenous decomposition  on a machine consists of 16 cores and 2 GPUs, the decomposed mesh is illustrated in Figure \ref{pic_hetroDecom} where the large subdomains at the left is associated to the GPUs. \\

\section{EVALUATION}
Our library is implemented in CUDA, C and C++ and plugged into OpenFOAM-ext1.6. The experimental platforms are: 
\begin{itemize}
\item A single node consisting of two NUMA sockets; each socket is connected to GPU device Tesla C2050. One socket has 6 cores Intel Xeon CPU X5670  @ 2.93GHz. 
\item A single node consisting of two NUMA sockets, which are connected to two GPU devices: GeForce GTX TITAN. One socket has 8 cores Intel Xeon CPU E5-2650 0 @ 2.00GHz. 
\end{itemize}
We used two test cases in our experiments:
\begin{itemize}
\item The lid-driven cavity flow test case contains the solution of a laminar, isothermal and incompressible flow within a three-dimensional cubic geometry using the icoFoam solver. The top boundary of the cube is a moving wall that moves in the \emph{$x$} direction, while the rest are static walls.
\item The heat equation with Dirichlet boundary condition test case is solved over a three-dimensional cubic geometry using laplacianFoam.
\end{itemize}

\begin{figure}[h!]
\begin{center}
\includegraphics[width=0.4\linewidth,height=0.27\textheight]{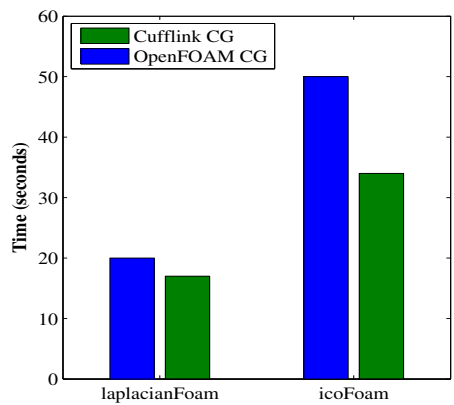} 
\caption{ Execution time of the selected OpenFOAM applications. Both internally call either OpenFOAM CG solver on CPU or Cufflink CG on GPU Tesla C2050 with convergence tolerance of $10^{-6}$, for a $100 \times 100 \times 100$ 3D domain.}
\label{tesla_base}
\end{center}
\end{figure}

\begin{figure}[h!]
\begin{center}
\includegraphics[width=0.4\linewidth,height=0.27\textheight]{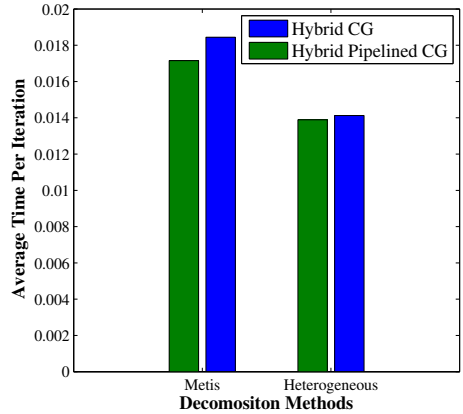} 
\caption{Heterogeneous decomposition compared against homogeneous MeTiS decomposition on a Tesla node, for a $100 \times 100 \times 100$ 3D domain. }
\label{heterodecom}
\end{center}
\end{figure}

Figure \ref{tesla_base} shows a comparison between OpenFOAM-CG and Cufflink-CG, which was running on a Tesla C2050. Cufflink is an open source library that extends the CG and BiCG solvers of OpenFOAM library to be executed on GPU. We measure the execution time per one step running of icoFoam and laplacianFoam as shown in the x-axis. We observe that Cufflink CG  was able to reduce the execution time by approximately 40\% when compared with OpenFOAM CG. \\

The heterogeneous decomposition is compared against homogeneous MeTiS decomposition on one Tesla node, as shown in Figure \ref{heterodecom}. This test uses the hybrid solvers, which solve a system of linear equations for a 7-point Laplacian in a 3D slab consisting of a million cells. The computational domain is decomposed into two subdomains. One subdomain is computed on the GPU Tesla C2050, whereas the other one on the CPU. Figure  \ref{heterodecom}  shows that the heterogeneous decomposition provides improvement of the average time per iteration around 20\%. The heterogeneous decomposition is based on empirical measurements, equation (\ref{percentage_f1}), so does not depend upon an intricate model. It is generalizable to many problems. However, the CFD applications may consist of different computational kernels, which require more auto-tuning using the heterogeneous decomposition. This step is noted for future work, and the current implementation is a stepping stone towards balancing computational kernels across highly heterogeneous devices. In this work, we try to optimize the most computational intensive phase in the application, and show how this optimization can accelerate the whole applications.  \\

\begin{figure}[h!]
\begin{center}
\emph{$a)$} \includegraphics[width=0.46\linewidth,height=0.28\textheight]{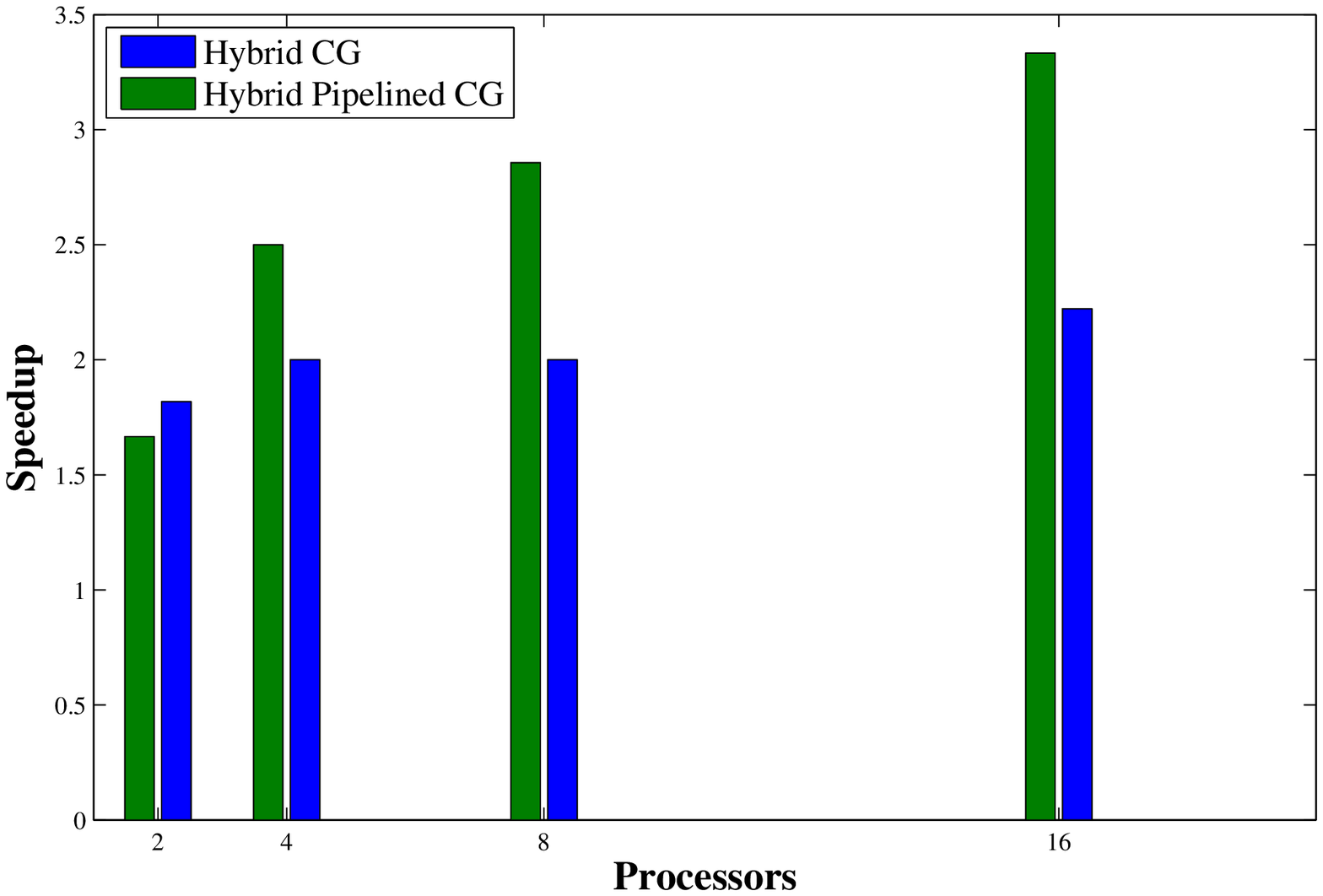}
\emph{$b)$} \includegraphics[width=0.46\linewidth,height=0.28\textheight]{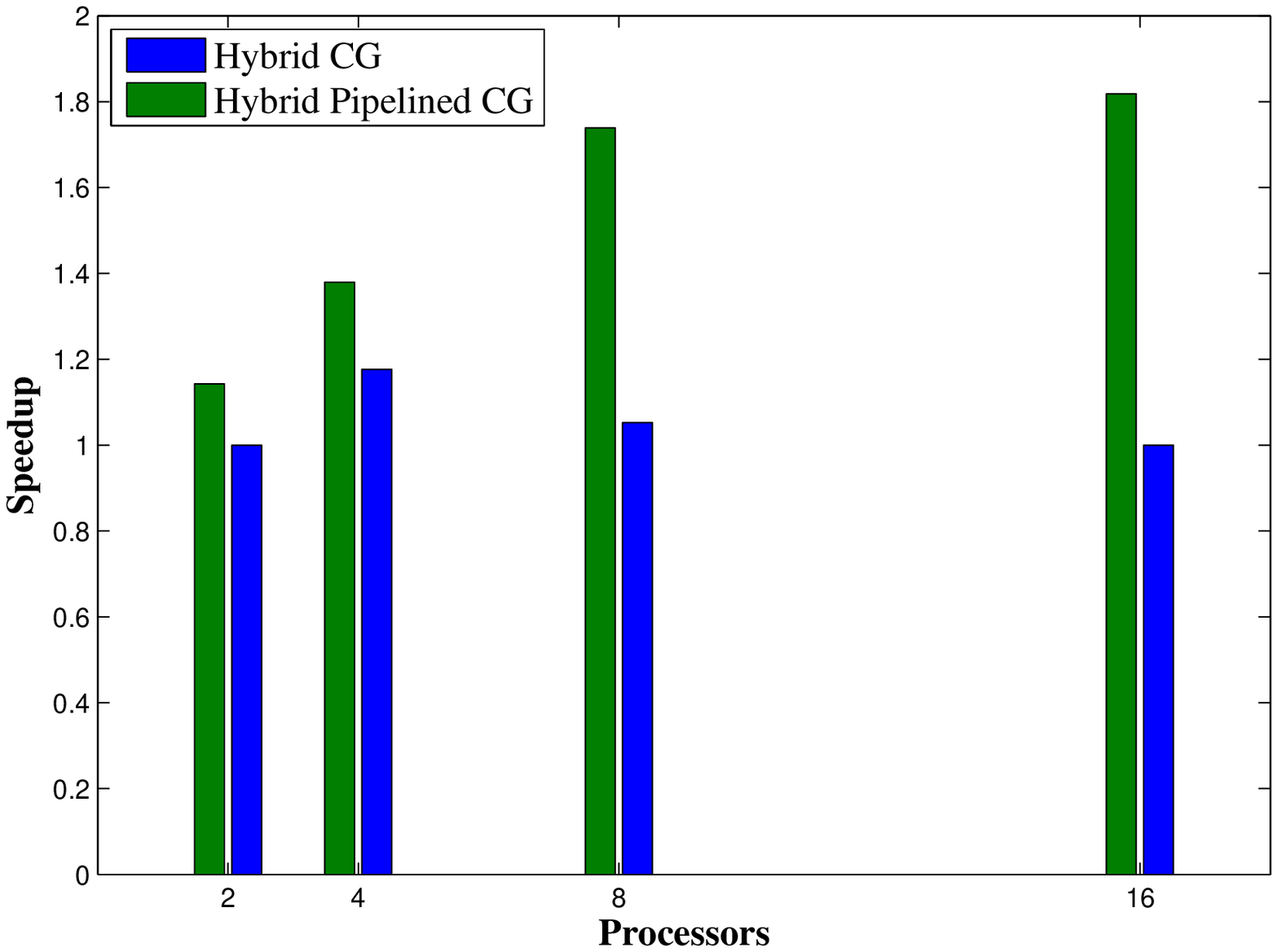}
\caption{Hybrid solvers strong scaling measured as speedup of the execution time relative of different parallel run over the icoFoam with a tolerance of $10^{-6}$ calling Cufflink-CG on a Tesla node. The mesh size in a) consists of 100,000 control volumes, and in b) of 1 million control volumes. }
\label{icospeedup}
\end{center}
\end{figure}  

\begin{figure}[h!]
\begin{center}
\emph{$a)$} \includegraphics[width=0.46\linewidth,height=0.28\textheight]{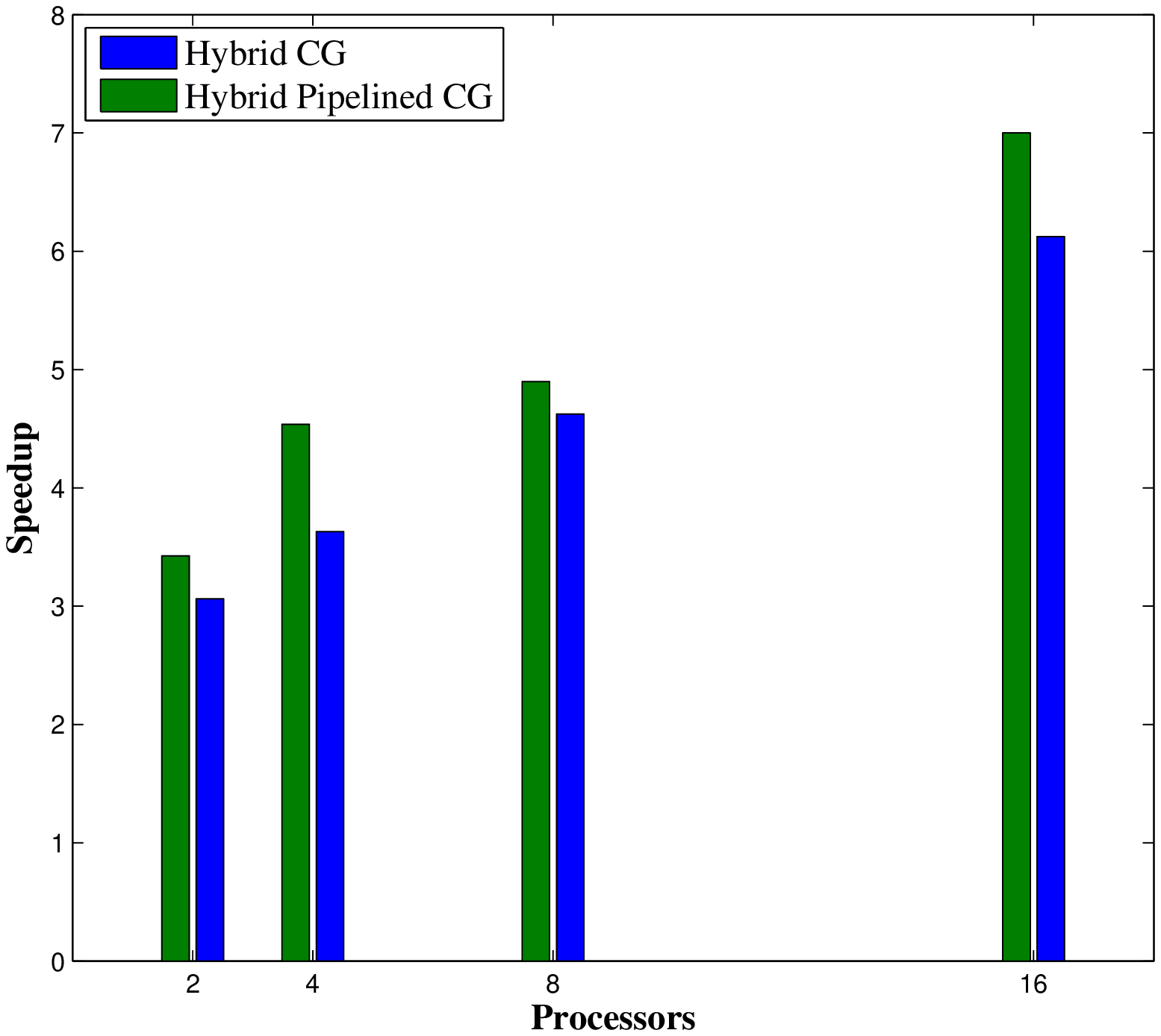}
\emph{$b)$} \includegraphics[width=0.46\linewidth,height=0.28\textheight]{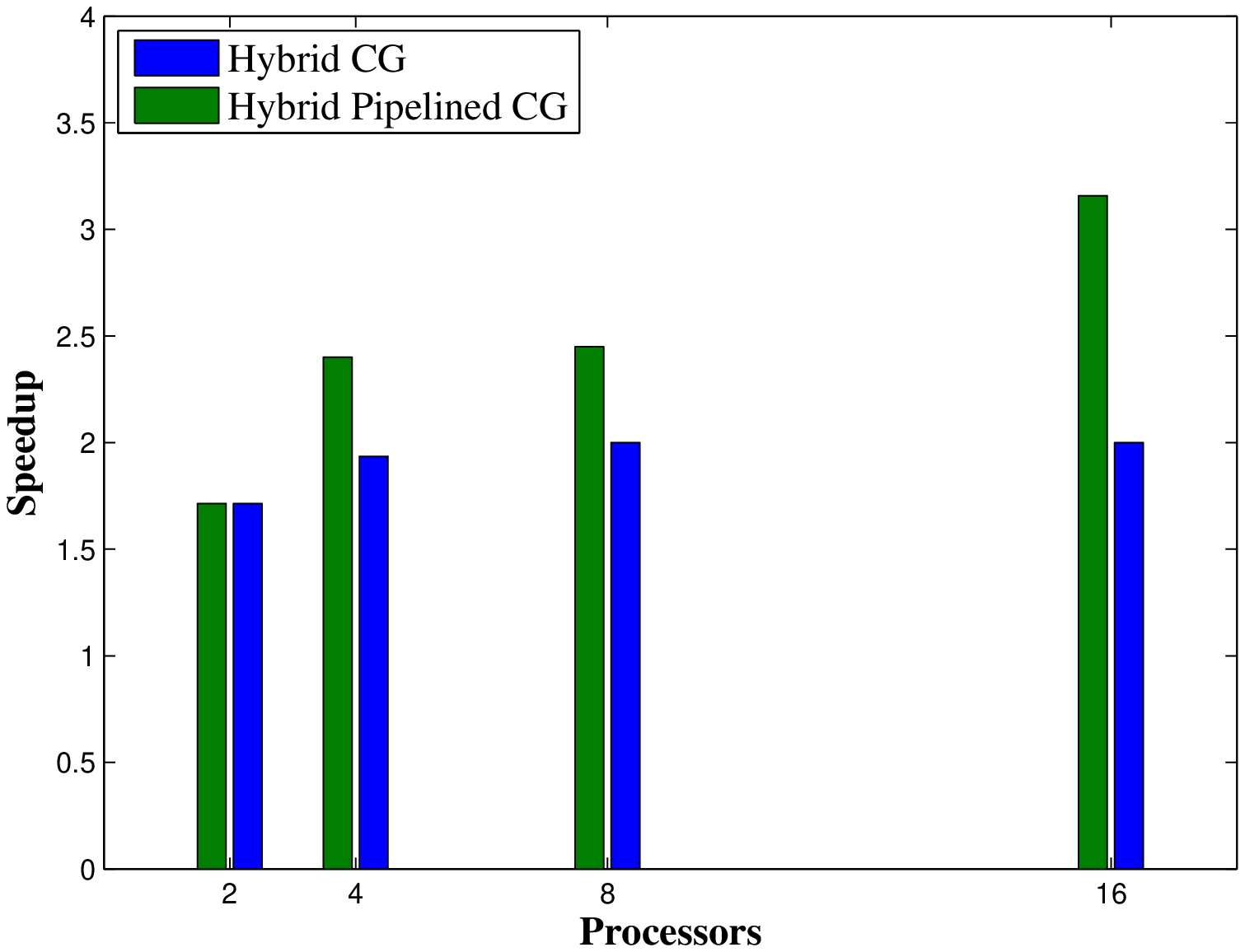}
\caption{Hybrid solvers strong scaling measured as speedup of the execution time relative of different parallel run over the laplacianFoam with a tolerance of $10^{-6}$ calling Cufflink CG on a Tesla node. The mesh size in a) consists of 100,000 control volumes, and in b) of 1 million control volumes.}
\label{lapspeedup}
\end{center}
\end{figure}  

We examine the speedup on the end-to-end computation of icoFoam and laplacianFoam on a Tesla node. Figure \ref{icospeedup}, shows the speedup of the total execution time of icoFoam using hybrid solvers over icoFoam using Cufflink CG on a GPU. The test case used to benchmark icoFoam is a 3D lid-driven cavity flow. It shows approximately 2x speedup on fixed problem size. The poor strong scalability is due to the limitation of the unoptimized stages of PISO algorithm. We only optimized the second stage, pressure solution, using hybrid solvers. Figure \ref{lapspeedup} shows the speedup of the total execution time of laplacianFoam using hybrid solvers over laplacianFoam using Cufflink GPU CG. The test case is a 3D heat equation. We observe a speedup about 7x in \emph{$a)$} and 3.5x in \emph{$b)$} better relative to Cufflink CG on a GPU. \\


\begin{figure}[h]
\begin{center}
\includegraphics[width=0.6\linewidth,height=0.27\textheight]{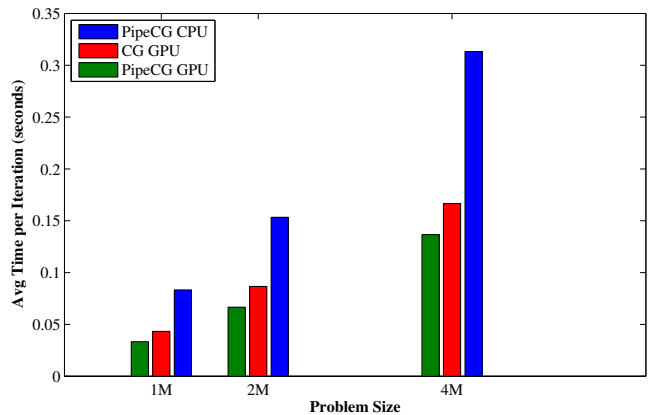}
\caption{Time per iteration comparison of pipelined CG MPI only, hybrid CG, and hybrid pipelined CG on different mesh sizes running on one socket, which is connected to a NVIDIA GeForce GTX TITAN GPU device, for a 7-point Laplacian in a 3D slab with a tolerance of $10^{-6}$.}
\label{titan_com}
\end{center}
\end{figure}

We also study the performance on an advanced GPU device, i.e., a Geforce GTX TITAN. In Figure \ref{titan_com}, we examine the performance of the CG solvers on TITAN, and compare three variants: pipelined CG on CPU (PipeCG CPU), CG on GPU (CG GPU), and pipelined CG on GPU (PipeCG GPU). While the execution time increases with mesh size, the rate of increase --- slope of graph --- for CPU solver is steeper than GPU solvers, indicating that both GPU solvers also achieves better performance. The performance of hybrid solvers is limited by device memory and PCIe bandwidth. The highest performance achieved is around 800 MFLOPs/s on Fermi and 2 GFLOPs/s on Titan, therefore, the current implementation is PCIe bandwidth limited according to the roofline model Figure \ref{perf1}. One of the main bottlenecks in achieving linear scalability and better efficiency, as shown in Figures \ref{perf1}, is the interface between the CUDA and the OpenFOAM structures. We suggest that OpenFOAM core classes such as \emph{lduMatrix} should be revised to allow for a storage of small, contiguous cache-friendly blocks of scalars, and thus potentially reduce the number of cache misses due to the random access patterns in the current data structures. \\

 \begin{table}[h]
  \begin{center}
     \caption{Arithmetic Intensity of Conjugate Gradient Kernels, where \emph{$N$}: no. of rows, \emph{$NNZ$}: number of non-zeros, and \emph{$N \simeq 7 NNZ$}.}
    \label{ai_cg}
    \begin{tabular}{ | l | l | l |  p{4cm} |}
    \hline
    Kernel & FLOPs & BYTEs & Arithmetic Intensity  \\ \hline
    Sparse Matrix-Vector Multiplication (CSR) & \emph{$14 N$} & \emph{$112 N$} & \emph{$0.125$} \\ \hline
    Vector-Vector Update & \emph{$2*N$} & \emph{$24*N$} & \emph{$0.083$} \\ \hline
    Vector Inner Product & \emph{$2*N$} & \emph{$16*N$} & \emph{$0.125$} \\\hline
    \end{tabular}
    \end{center}
    \end{table}

As mentioned previously, the performance of the CG algorithm per loop depends on the performance of each individual kernel. The arithmetic intensity of these kernels are tabulated in Table \ref{ai_cg}. Such low arithmetic intensity demands high memory-bandwidth, and thus many flops are lost waiting for data. Given that the operations in the CG method are memory bound, the best performance is reached when the memory throughput is maximized. Combining the vector operations into larger kernel allows better hiding of memory latency. There are two ways to optimize the vector operations on GPUs: first by processing multiple vectors that will initiate more memory transactions, and thus the smaller the vector size the better the performance. The other way is processing one vector of large size that is, apparently, the only way to maximize the performance of the reduction operation; because it can not be combined with other operations. Therefore, the vector reduction is a bottleneck. In addition, it requires data transfer over PCIe between host and device, which is an order of magnitude slower than the bandwidth of the memory bus on GPU. These data movements also appear in the sparse matrix-vector multiplication while updating the interfaces among neighboring subdomains. These communications negatively and significantly affect the achievable performance.     
~\\

%

\begin{figure}[h!]
\begin{center}
\emph{$A)$} \includegraphics[width=0.47\linewidth,height=0.3\textheight]{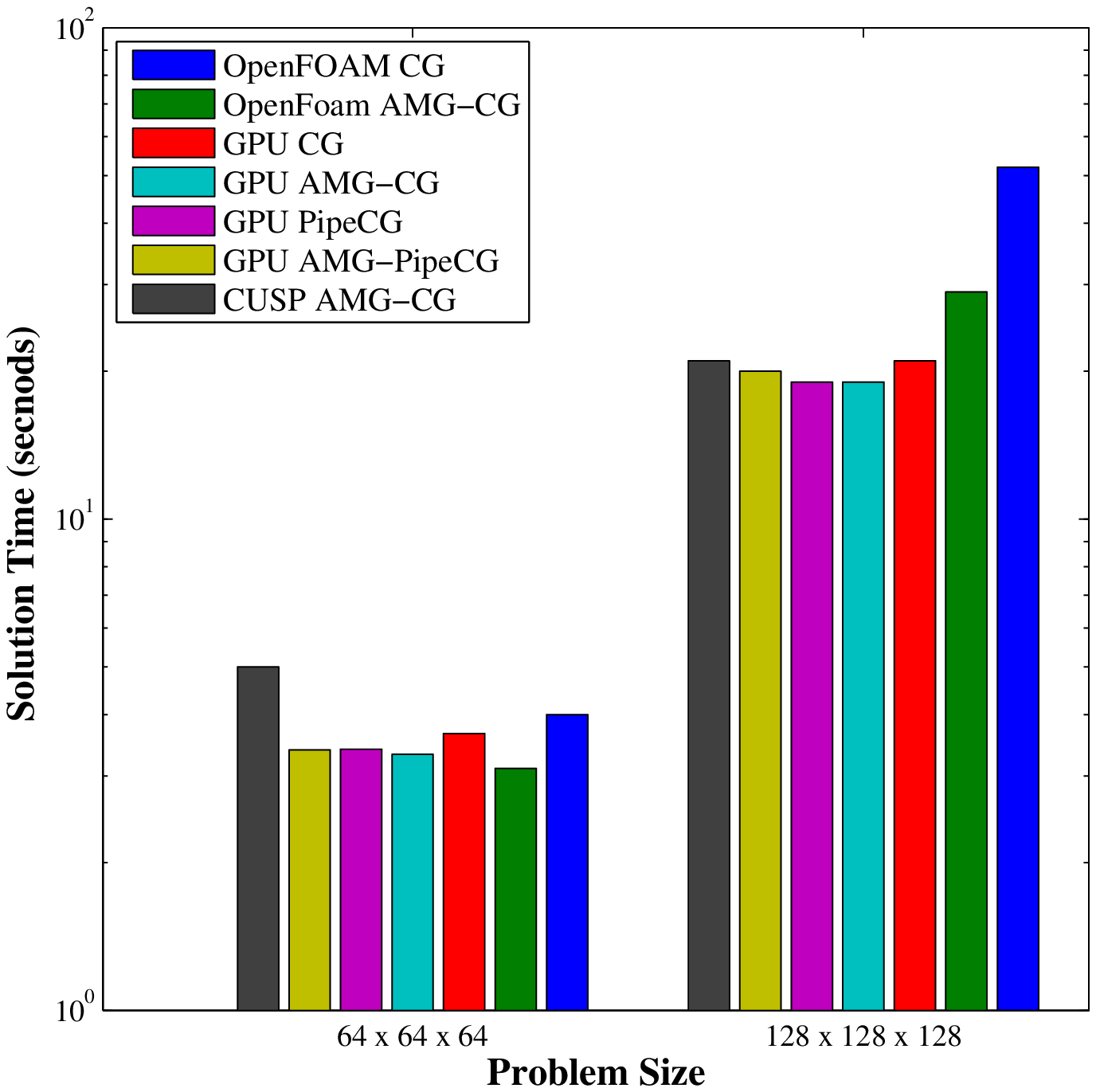}
\emph{$B)$} \includegraphics[width=0.47\linewidth,height=0.3\textheight]{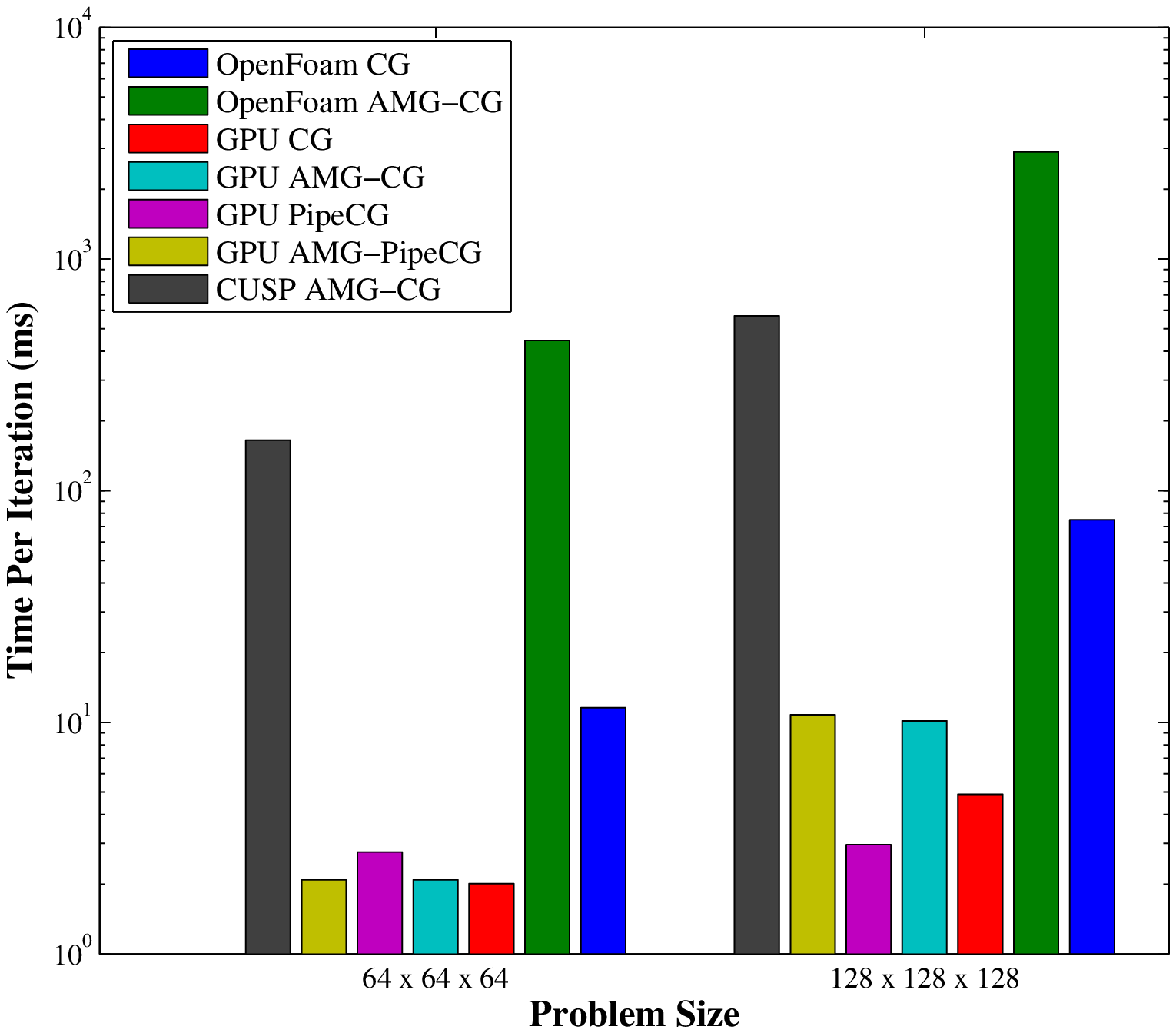}
\caption{Execution time and time per iteration of a steady state 7-point Laplacian with Drichlet boundary conditions in a 3D cube solved using different none/preconditioned CG solvers with a tolerance of $10^{-8}$ on different mesh sizes running on one socket, which is connected to a NVIDIA GeForce GTX TITAN GPU device.}
\label{pcgs1}
\end{center}
\end{figure} 

 In Figure \ref{pcgs1}, we apply an aggregation-based algebraic multigrid  preconditioner \cite{cusp_amg} to hybrid CG and hybrid Pipelnied CG solvers and compare it to OpenFOAM (CPU) and CUSP (GPU). The AMG preconditioner is implemented in the CUSP library. It performs a V-cycle of an aggregation-based AMG with weighted Jacobi iterations with $\omega=\frac{4}{3}$ as a smoother within multigrid \cite{cusp_amg, cusp}. In OpenFOAM , an agglomeration-based algebraic multigrid preconditioner is applied to its CG solver that performs a V-cycle of an agglomeration-based AMG with diagonal Incomplete-Cholesky smoother. Our hybrid solvers are preconditioned using an aggregation-based AMG included within CUSP library. The increase in the number of iterations required to converge by our solvers on GPU compared to OpenFOAM AMG-CG, which is performed on a CPU sequentially, indicates the effects of the following factors: irregularly shaped aggregates generated by the parallel aggregation method based on a distance-2 maximal-independent-set within AMG \cite{cusp_amg}, the smoother, and the multigrid hierarchies. Despite that, our solvers are faster than OpenFoam solver, as shown in Figure \ref{pcgs1}, also GPU AMG-CG performs better than CUSP AMG-CG indicating a better implementation of CG.     
 ~\\ 
 
\begin{table}[h]
\begin{center}
\caption{Number of iterations to solve a steady state 7-point Laplacian with Drichlet boundary conditions in a 3D cube using different none/preconditioned CG solvers with a tolerance of $10^{-8}$ on different mesh sizes running on one socket, which is connected to a NVIDIA GeForce GTX TITAN GPU device.}
\label{pcgs2}
\begin{tabular}{|
>{\columncolor[HTML]{FFFFFF}}l |
>{\columncolor[HTML]{FFFFFF}}l |
>{\columncolor[HTML]{FFFFFF}}l |l|
>{\columncolor[HTML]{FFFFFF}}l |
>{\columncolor[HTML]{FFFFFF}}l |
>{\columncolor[HTML]{FFFFFF}}l |}
\cline{1-3} \cline{5-7}
Mesh Size $N^{3}$ & N=64 & N=128 &  & {\color[HTML]{333333} Mesh Size $N^{3}$} & {\color[HTML]{333333} N=64} & {\color[HTML]{333333} N=128} \\ \cline{1-3} \cline{5-7} 
OpenFOAM CG & 346 & 695 &  & {\color[HTML]{333333} OpenFOAM AMG-CG} & {\color[HTML]{333333} 7} & {\color[HTML]{333333} 10} \\ \cline{1-3} \cline{5-7} 
GPU CG & 325 & 636 &  & {\color[HTML]{333333} GPU AMG-CG} & {\color[HTML]{333333} 28} & {\color[HTML]{333333} 32} \\ \cline{1-3} \cline{5-7} 
GPU Pipe CG & 157 & 297 &  & {\color[HTML]{333333} GPU AMG-PipeCG} & {\color[HTML]{333333} 28} & {\color[HTML]{333333} 32} \\ \cline{1-3} \cline{5-7} 
- & - & - &  & {\color[HTML]{333333} CUSP AMG-CG} & {\color[HTML]{333333} 32} & {\color[HTML]{333333} 37} \\ \cline{1-3} \cline{5-7} 
\end{tabular}
\end{center}
\end{table}
 

\section{CONCLUSIONS}
Memory bound applications, such as the OpenFOAM solvers, can take better advantage of the full hardware potential, which is now complex, hybrid and heterogeneous, if all resources are taken into account in a holistic approach. While many questions of ultimately attainable per node performance and multi-node scaling remain, the experimental results show that the hybrid implementation of both solvers outperforms state-of-the-art implementations of a widely used open source package. This work can be seen as a stepping stone towards optimized hybrid heterogenous CFD applications. The current limitations of the hybrid solvers and the heterogenous decomposition can be solved by more auto-tuning and dynamic load balancing scheduling, which adaptively balances the workload of different computational kernels during the run-time, by memory-aware work stealing.






\bibliographystyle{elsarticle-num}
\bibliography{mybibtexfile}






\end{document}